# Targeted material doping to enable the formation and coherent propagation of Laser Induced Periodic Surface Structures (LIPSS): the ZnO paradigm.

*Gonzalo Gomez-Munoz[1*], Rocio Ariza[1,†], Juan Francisco Ramos-Justicia[2], Fatima Cabello-Pardos[1], Antonio Pablo Pérez-Marín[1], Jose Gonzalo[1], Belén Sotillo[2], Binbing Xia[3], Paloma Fernandez[2], Javier Solis[1*]*

[1]Instituto de Optica-CSIC (IO-CSIC), c/Serrano 121, ES-28006 Madrid, Spain.
[2]Departamento de Física de Materiales - UCM, Pl. de las Ciencias 1, ES-28040 Madrid, Spain.
[3]Institute des Nanosciences de Paris (INSP), Sorbonne Univ-CNRS, 4 place Jussieu,75005-Paris, France.

* Corresponding authors: gonzalo.gomez@io.cfmac.csic.es, j.solis@io.cfmac.csic.es
† Present address: Centre of Membrane Separation, Adsorptions, Catalysis and Spectroscopy for Sustainable Solutions – KU Leuven, Celestijnenlaan 200F, BE-3001 Leuven, Belgium

## Abstract

This work explores how the initial free carrier density governs the formation and coherent propagation of Low Spatial Frequency Laser-Induced Periodic Surface Structures (LSF-LIPSS). As a paradigm we have analyzed the behavior of ad-hoc-doped ZnO films. ZnO is a multifunctional Transparent Conducting Oxide (TCO) with many scientific and technologic uses including applications in biomedicine, photonics, sensing or photocatalysis. Many of them would benefit from the use of large area, electrically anisotropic surfaces. By introducing small amounts of $Yb^{3+}$ as donor dopant in ZnO films grown by Pulse Laser Deposition (PLD), we have substantially improved its carrier density at room temperature (RT) and its electrical conductivity while preserving the functional properties of pure ZnO. For RT carrier densities $n_e > 10^{19}$ cm$^{-3}$, coherent propagation of LSF-LIPSS has been successfully achieved upon irradiation with 350 fs laser pulses at 1030 nm at high repetition rate (500 kHz) and for laser scanning speeds well above 1 m/s. However, the ablation mechanism (ZnO sublimation at the LIPSS valleys) and the large melting temperature of ZnO, make the thermal conductivity of the substrate, the film thickness and the irradiation pulse duration to play a crucial role in determining the morphology of the LSF-LIPPS propagated over large areas, especially in avoiding film cracking. The results achieved upon optimization fully confirm the role of free carrier density in the large-scale propagation of high quality LSF-LIPSS in doped ZnO and enable using these structures to easily generate large electrical anisotropies over macroscopic regions with conductivities along the LIPSS several times larger than transversally.

## 1. Introduction

Recently, the formation and coherent propagation of Low Spatial Frequency (LSF) Laser-Induced Periodic Surface Structures (LIPSS) [1,2] over large areas have been successfully achieved in two different transparent conductive oxides (TCOs), ITO [3] and FTO [4] upon fs-laser pulsed irradiation at 1030 nm with laser scanning speeds well above 1 *m*/*s*. The control in the formation of the LSF-LIPSS structures with the processing parameters leads to the production of highly anisotropic conductive surfaces that can be used in several applications like transparent heaters [4,5], sensors [6], templates to produce aligned metal nanostructures [7] or substrates for electrically controllable wettability [8]. However, ITO and FTO are very different from the compositional point of view. Indeed, the key parameter controlling the evolution of micro-scale morphology during the LSF-LIPSS formation process is the preferential loss of In in the case of ITO [3] or the F/Sn enrichment in the film upon multiple pulse exposure in the case of FTO [4]. The common characteristic to explain their facility to generate coherently propagated LSF-LIPSS over extremely large areas seems not thus to be related to their chemical composition but to the way laser energy is coupled into the material. At the near-infrared (NIR) laser wavelength used (1030 nm), both materials are nearly transparent, but they show a relatively high free carrier density (~ $10^{20}$ $cm^{-3}$) that facilitates avalanche ionization upon laser exposure [4] and the subsequent formation of LIPSS upon multiple pulse exposure.

To investigate the correlation between the free carrier density and the formation and coherent propagation of LIPSS, in this work we have chosen a very different material in terms of chemical composition such as ZnO. The material has been ad-hoc doped to modify its free carrier density at room temperature (RT) in a controlled manner, and then laser processed to investigate the conditions for LIPSS formation and propagation. In addition, ZnO is a very promising material for many technological applications.

ZnO is a wide band-gap semiconductor that has received considerable attention over many years due to its multifunctional characteristics [9,10]. It has a relatively large bandgap ($E_g$ ~ 3.37 eV) and a large exciton binding energy of 60 meV, potentially enabling efficient room-temperature exciton-based emitters, and sharp transitions facilitating very low threshold semiconductor lasers [11]. Still, self-compensation and other effects [12] (low solubility of acceptor impurities and deep acceptor energy levels) hinder the obtention of *p*-type ZnO. The material is thus normally *n*-type [12] with a relatively high carrier density at room temperature related to the presence of different types of defects depending on the synthesis approach used [13,14]. Indeed, heavy doping with trivalent elements enables to achieve carrier densities around and above $10^{20}$ $cm^{-3}$ in doped-ZnO. Together with the inherent transparency of ZnO in the VIS-NIR region, this kind of doping makes Group-III-doped ZnO films (AZO, GZO…) to be especially useful for transparent electronics applications [15]. This includes applications as conductive electrode in light absorbing and emitting devices [16], solar cells [17,18], as well as NIR plasmonics [19]. Other applications of ZnO due to its especial properties, like the generation of strong oxidizing species upon illumination, include its use in solar photocatalysis [20–22] and antibacterial [23,24] applications. Sensing [6,10] and electrowetting [8] applications of ZnO are also remarkable while the piezoelectric properties of ZnO have been used for the development of piezotronics and piezo-phototronics devices [25–27]. Many of these applications would benefit from the use of micro- nano-structured surfaces over large areas.

Among other structuring approaches, the use of LIPSS [1,2] offers a number of advantages, including an increase of the specific surface of the material (due to the micro- and nano topographic modulation) and the controlled modification of several surface properties of the material at high processing speed. LIPSS have enabled the development of applications involving the modification or improvement of the nucleation [28,29], wetting [30,31], tribological [32,33], optical [34,35], or sensing [6,36] properties of the material surface and very especially in the field of biomimetics [37].

This work explores how the initial free carrier density controls the formation and propagation of LSF-LIPSS for producing large-scale, laser-manufacturing applications in TCO´s. As a paradigm we have analyzed the formation and coherent propagation of LSF-LIPSS in ad-hoc doped ZnO films. Films produced by Pulsed Laser Deposition (PLD) ZnO have been doped with different amounts of $Yb^{3+}$ to modify their room temperature free carrier densities. The material response upon fs-laser processing at 1030 nm with 500 kHz fs-laser pulses has been subsequently analyzed. For doping levels leading to $n_e > 10^{19}$ $cm^{-3}$ formation and coherent propagation of LSF-LIPSS has been successfully achieved upon irradiation at processing speeds well above 1 m/s. In addition, as expected, LIPSS processed surfaces show a strong electrical anisotropy, with conductivities along the LIPSS about hundred times larger than transversally.

## 2. Experimental details

### 2.1. Pulsed Laser Deposited ZnO: $Yb^{3+}$ films

Yb doped ZnO thin films were deposited by PLD using an ArF excimer laser (λ=193 nm, 20 Hz repetition rate) in a high vacuum (HV) chamber. The laser beam was focused at angle of incidence of 45º on the surface of $Yb_2O_3$ doped ZnO sintered targets to lead to a laser fluence of 4.0 J/cm$^2$. The nominal $Yb_2O_3$ concentrations were 0.0, 1.0, 2.5, 5.0 and 10.0 wt%. The PLD targets were sintered from ZnO (Sigma-Aldrich, purity 99.9%, #MKCL7155), and $Yb_2O_3$ powders pressed at 2 T and heated at 1200 °C for 5 hours. The vacuum chamber was evacuated to a residual pressure of $10^{-6}$ mbar and then filled with a dynamic oxygen pressure of $0.7\text{-}2\cdot 10^{-2}$ mbar. The target was rotated at a few rpm during the deposition process to avoid fast-track formation on their surface. The target-substrate distance was 40 mm. The films were deposited on (100) p-type Si for calibration of the system. Then, the samples were prepared depositing the desired film on soda-lime glass (SLG), fused silica glass (FSG) and c-cut sapphire (Sapp.) substrates held at room temperature (RT), which were ultrasonically cleaned in acetone and ethanol before deposition.

### 2.2 Laser processing details

Femtosecond laser processing of the films was performed in ambient air using an Yb-doped amplified fiber laser (Satsuma HP$^2$, Amplitude Systemes) operating at a central wavelength of 1030 nm with a repetition rate of 500 kHz and a pulse duration of the system set to 350 fs using the system compressor alignment. The laser beam was conditioned using several elements to define its polarization and energy before reaching a ScanCube 14 galvo-scanner (Scanlab) equipped with an F-theta lens ($f$ = 100 mm), used to both, focus the beam and deflect it to scan the surface of the irradiation targets. The scanner provides a flat field, focusing area of 7 × 7 cm$^2$ and was operated with commercial software (LaserDESK, Scanlab) to control the scanning parameters. Typically, beam scanning was performed along the horizontal direction (parallel to the floor plane) with the beam polarization oriented vertically at the sample (perpendicular to the scanning direction). The spot diameter of the Gaussian-shaped laser beam at the sample surface was determined following the method proposed by Liu [38] to be $2\omega_0$ = 29.2 µm (1/e$^2$ intensity), where $\omega_0$ is the spot radius. The average power of the laser beam ($\bar{P}$) was determined with an Ophir 3A-PF-12 thermopile sensor with a sensitivity of 15 µJ and a 0.2% accuracy. For a given average power of the laser, the energy of the individual pulses ($E$) was:

$$E = \bar{P}/f_{rep}, \qquad [1]$$

being $f_{rep}$ the repetition rate of the laser. The local fluence pulse ($F$) at the sample site for a given pulse energy was:

$$F = 2\cdot E/(\pi\cdot\omega_0^2). \qquad [2]$$

### 2.3 Sample characterization

Optical transmission (T) spectra of the samples were measured using a Varian Cary 5000 UV-VIS-NIR spectrophotometer with a resolution of 5 nm over the 350-2500 nm range. Morphology was assessed by Optical Microscopy (OM) and Scanning Electron Microscopy (SEM) in Secondary Electron detection mode (SE). OM characterization was performed with a Nikon Eclipse-Ti Microscope using illumination at 460 nm. The top-view images of the LIPSS structures were Fast Fourier Transform (FFT) analyzed to quantify the LIPSS period, (Λ). For the SE-SEM characterization, a FEI Inspect-S SEM and a Jeol JEM 6500F FEG-SEM (equipped with an Oxford INCA Energy 200 microanalysis system) were used to get top-view, high magnification images of the induced structures.

The composition of as-deposited films was determined from Ion Beam Analysis (IBA) techniques. Cation contents were determined by Rutherford Backscattering Spectrometry (RBS) using 80 nA $^4He^+$ beams of 1.9 MeV, in a 1mm diameter beam spot, with particle detection at 165° scattering angle. Zn ([Zn]) and Yb ([Y]) areal densities were obtained from the RBS spectra by directly comparing their respective signals with the absolute values obtained from the Bi reference ([Bi]= 5.66±0.17 × $10^{15}$ at/ $cm^2$), using the formula [3]:

$$\frac{[X]}{[Bi]} = \frac{N_X}{N_{Bi}} \times \left(\frac{Z_{Bi}}{Z_X}\right)^2 \times \frac{Q_{Bi}}{Q_X} \qquad [3]$$

where X= Zn, Yb and Ni, Zi and Qi (with i= X, Bi) are respectively the number of counts corresponding to element i, the atomic number of element i, and the total incident beam charge for the films and Bi reference. In addition, Nuclear Reaction Analysis (NRA) was used to determine the $^{16}O$ ([$^{16}O$]) areal density, using the $^{16}O(d, p)^{17}O$ nuclear reaction [**]. Absolute values were obtained within 3% by comparison with a $Ta_2O_5$ reference with a known [$^{16}O$] = 6,97 × $10^{17}$ at/$cm^2$. An 860 keV deuteron beam of 85 nA in a 2 mm beamspot was used, and the protons produced by the nuclear reactions were detected at 150° in a 300 $mm^2$ detector behind a 10 μm mylar foil which stopped the flux of elastically scattered deuterons from saturating the detector. The evolution of the Yb/Zn (at.%) in the laser processed regions was assessed by EDX measurements at the Jeol JEM 6500F FEG-SEM using a 15 kV excitation voltage.

The macroscopic electrical properties were assessed in large, squared areas (6 × 6 $mm^2$) that were electrically isolated from the rest of the film by inscribing a deep laser ablated line at high power after processing. The electrical transport characterization of the pristine samples was performed at RT in the van der Pauw configuration by means of sheet resistance and Hall effect measurements in an ECOPIA HMS-7000. In absence of magnetic field, the system introduces current between two of the probes and measures the induced voltage between the other two, using this to directly calculate the sheet resistance ($R_\square$) and the resistivity ($\rho$) of the sample. Under an applied magnetic field of B = 0.556 T in RT measuring mode, the Hall effect measurements allow to obtain several electronical properties of the samples, such as the density ($n_e$) and mobility ($\mu_e$) of their carriers.

To quantify the electrical anisotropy of the processed areas, sheet resistance measurements were performed using a 4-Point-Probe Ossila$^{TM}$ device following the criterium stablished in [4]. To avoid an overestimation of the sheet resistance measured, the system introduces a correction factor based on the geometry of the measured sample (which is manually introduced in the control software of the equipment). The probe spacing used was 1.27 mm, and the target current for the measurements was adjusted when needed to ensure the measurement validity considering the ranges given by the equipment manufacturer. The applied voltage was increased in steps of 100 mV until reaching a maximum value of 10 V. During the measurement, depending on how the tips are placed, parallel or transversal to the LIPSS orientation, we have determined two different values of sheet resistance, namely $R_{\square||}$ and $R_{\square\perp}$. The electrical anisotropy factor ($\xi$) was then calculated as

$$\xi = R_{\square\perp} / R_{\square||}. \qquad [4]$$

## 3. Results and discussion

### 3.1. Pulsed Laser Deposited ZnO:$Yb^{3+}$ films

Sample synthesis was carried out in two steps. In the first one we used the optimum experimental conditions previously determined in our experimental system for producing high quality undoped ZnO films by Pulsed Laser Deposition (PLD) at RT [39]: an $O_2$ dynamic pressure of $2.0 \times 10^{-2}$ mbar and a laser fluence of $\sim$ 4 J/cm$^2$. The aim was to achieve undoped ZnO films of good quality and then use these PLD settings to grow doped films from targets with different $Yb_2O_3$ content.

In this initial step, after reaching a background pressure of $10^{-6}$ mbar, PLD was performed under an oxygen pressure of $2.0 \times 10^{-2}$ mbar. The carrier densities obtained in films from the different targets were then measured and the films irradiated to generate LSF-LIPSS. In the second optimization step, the $O_2$ pressure in the PLD chamber was optimized to increase further the free carrier density of the films.

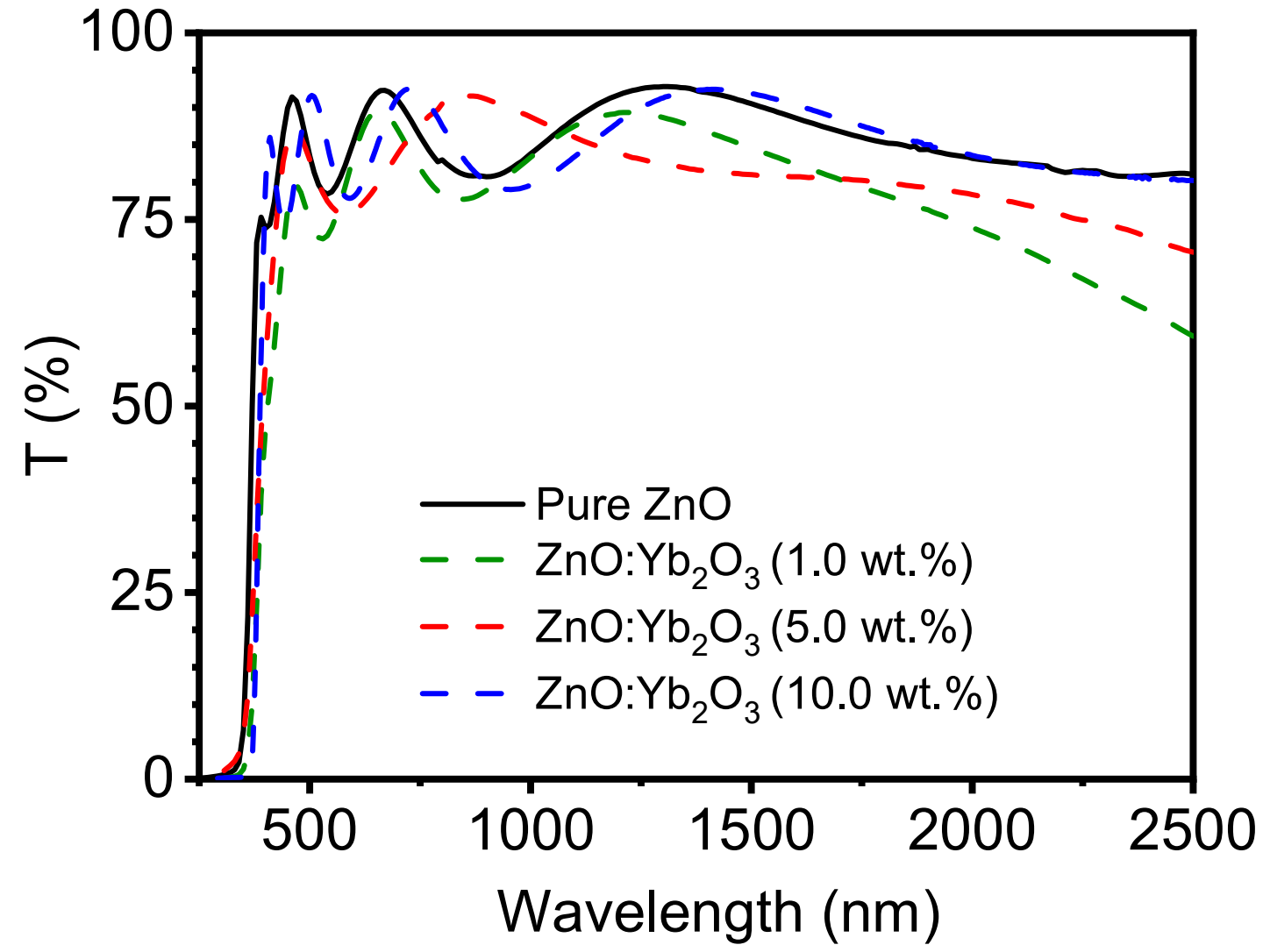

**Figure 1.** Optical transmission spectra of films grown from targets with different contents of ZnO:$Yb_2O_3$ on glass substrates. The film thickness was ~ 350 nm in all cases.

**Figure 1** shows the optical transmission spectra corresponding to films grown on glass from targets with different $Yb_2O_3$ content (0 – 10% wt.%) grown at 2.0 × $10^{-2}$ mbar $O_2$ pressure in the chamber. Pure ZnO films target were very transparent over the whole measured spectral interval (250 – 2500 nm) with a maximum transmission close to 90%. The transparency of the films is evident in the constant amplitude of the transmission oscillations for increasing wavelengths. This is consistent with a relatively high density, a low level of oxygen vacancies and a relatively good crystalline quality even when the films were grown at RT [40]. The measured index of refraction of the films was $n$ = 2.18 at 550 nm which is in the range of values typically observed in ZnO films. The measured free electron density of $n_e = 8.8 \times 10^{18}$ $cm^{-3}$ and sheet resistance of 12.2 kΩ/□ values evidence the films are not free from point defects (mainly oxygen vacancies), as expected from self-compensation and other mechanisms characteristic of ZnO [15].

The incorporation of $Yb^{3+}$ in the films induces a modification of the refractive index as it is shown in **Figure 1** (for instance for a 5.0 at% doping, the refractive index $n$ (550 nm) = 2.07) with a shift in the transmission oscillations, and a transmission decrease for wavelengths beyond ~ 1 µm. This behavior is caused by the free electrons generated by doping. The most relevant electrical parameters of the films grown at 2 × $10^{-2}$ mbar are shown in **Table 1**. The table also includes the ratio of Yb/Zn atoms both in the target and in the films, as determined by RBS measurements (see the Supporting Information File, SIF, for further details).

The free carrier density increases by one order of magnitude (from $n_e = 8.76 \times 10^{18}$ cm$^{-3}$ to $n_e = 9.13 \times 10^{19}$ cm$^{-3}$) for a doping level of just 2.5 wt.% in the target. Nevertheless, beyond 2.5 wt.% the carrier density and the carrier mobility decrease, and the resistance of the films increases again.

The other effect that can be appreciated in **Table 1** is the enrichment in Yb content of the films when compared to that of the target. This effect is expected due to the much higher mass of the Yb species in the laser generated plasma plume which are less scattered by the oxygen atoms present in the chamber than the Zn species [41]. To facilitate the discussion, from now on we will refer to the films in terms of the target composition they derive from.

**Figure 2** shows SEM images of films having different $Yb_2O_3$ contents. For the undoped films and doping levels up to 2.5 wt.%, the grain size is in the 100-200 nm range while a further increase of amount of $Yb_2O_3$ leads to a substantial decrease of the grain size to values in 20 – 40 nm range. In some regions, see **Figure 2(b)**, it is possible to find a few droplets caused by splashing of the liquid phase during the PLD process. Still, the use of an ablation wavelength (193 nm) well below the bandgap of ZnO enables the production smooth, dense and optically conductive films of ZnO [40]. All the observed morphological features are consistent with the observations in previous works [15,40,42].

**Table 1.** Experimental Yb/Zn atoms ratio in the PLD targets (nominal) and in the films (as determined by RBS measurements). The most relevant electrical parameters measured in 335 nm-thick films deposited on soda-lime glass substrates at $2 \times 10^{-2}$ mbar are also included.

| $Yb_2O_3$ [wt.%] | Yb/Zn [at.%] (target / film) | Carrier density $n_e$ [cm$^{-3}$] | Carrier mobility [cm$^2$/(s·V)] | Resistivity [Ω × cm] | Sheet resistance [Ω / □] |
|---|---|---|---|---|---|
| 0.0 | 0.00 / 0.00 | $8.76 \times 10^{18}$ | 1,06 | $3.96 \times 10^{-1}$ | $12.2 \times 10^{3}$ |
| 1.0 | 0.42 / 0.48 | $3.43 \times 10^{19}$ | 5.04 | $1.85 \times 10^{-2}$ | $5.79 \times 10^{2}$ |
| 2.5 | 1.05 / NA | $9.13 \times 10^{19}$ | 5.89 | $6.04 \times 10^{-2}$ | $1.84 \times 10^{2}$ |
| 5.0 | 2.17 / 3.45 | $3.0 \times 10^{19}$ | 1.52 | $1.3 \times 10^{-1}$ | $3.92 \times 10^{3}$ |
| 10.0 | 4.58 / 5.93 | $9.3 \times 10^{18}$ | 0.4 | $3.5 \times 10^{-1}$ | $1.00 \times 10^{4}$ |

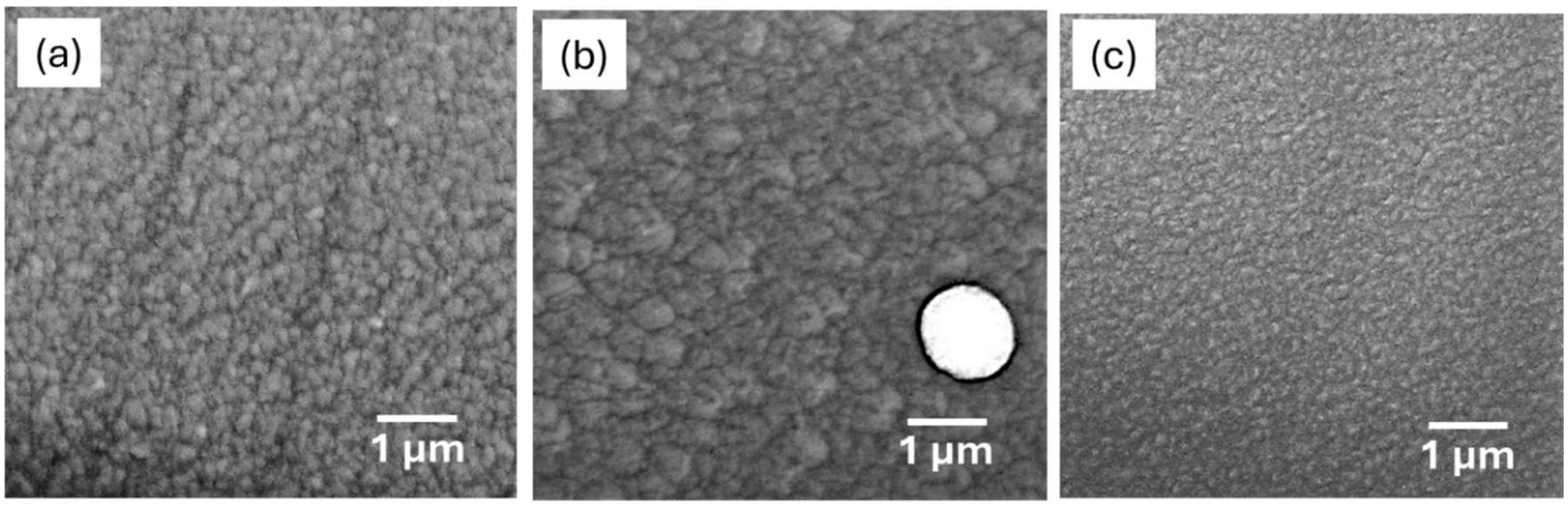

**Figure 2.** SEM micrographs from films with different nominal compositions. (a) $ZnO:Yb_2O_3$ (1.0 wt.%), (b) $ZnO:Yb_2O_3$ (2.5 wt.%), and (c) $ZnO:Yb_2O_3$ (10.0 wt.%) deposited on glass substrates. The film thickness in all cases corresponds to ~ 350 nm.

Along with the indicated changes in the grain size, Yb doping also strongly affects crystalline quality and structure of the material due to the larger ionic radius of $Yb^{3+}$ (~ 0.87 Å) compared to $Zn^{2+}$(0.74 Å). This can be clearly appreciated in the evolution of the X-ray diffractograms for the different film compositions that are shown **Figure 3**. Pure ZnO film shows the characteristic (0 0 2) reflection of hexagonal ZnO with a clearly visible (1 0 3) reflection. This indicates that the film is textured along the direction perpendicular to the substrate. For the 1.0 wt.% and 2.5 wt.% films, the lattice distortion caused by $Yb^{3+}$ ions is first evidenced by a clear diminution of the (0 0 2) peak and the increase of the (1 1 0) peak. At the same time the (1 0 1) peak, barely visible in the pure ZnO film, increases with the doping level and the symmetry-forbidden (1 0 0) reflection appears, indicating a strong deformation of the lattice. Such a deformation and decrease of grain size is even clearer for the highest doping level analyzed (10.0 wt.%) for which the (0 0 2) reflection nearly vanishes and the 30 – 40 degrees region in the 2θ plot is dominated by the (1 0 0) and (1 0 1) peaks. These features also suggest a strong diminution of the film texture along the [0 0 2] direction (perpendicular to the substrate) in the doped films. This can be well appreciated in the evolution of the ratios of the peak intensities ($I_{(0\,0\,2)}/I_{(1\,0\,0)}$) and ($I_{(0\,0\,2)}/I_{(1\,0\,1)}$) as a function of the doping that are included in SIF file (**Figure S2**).

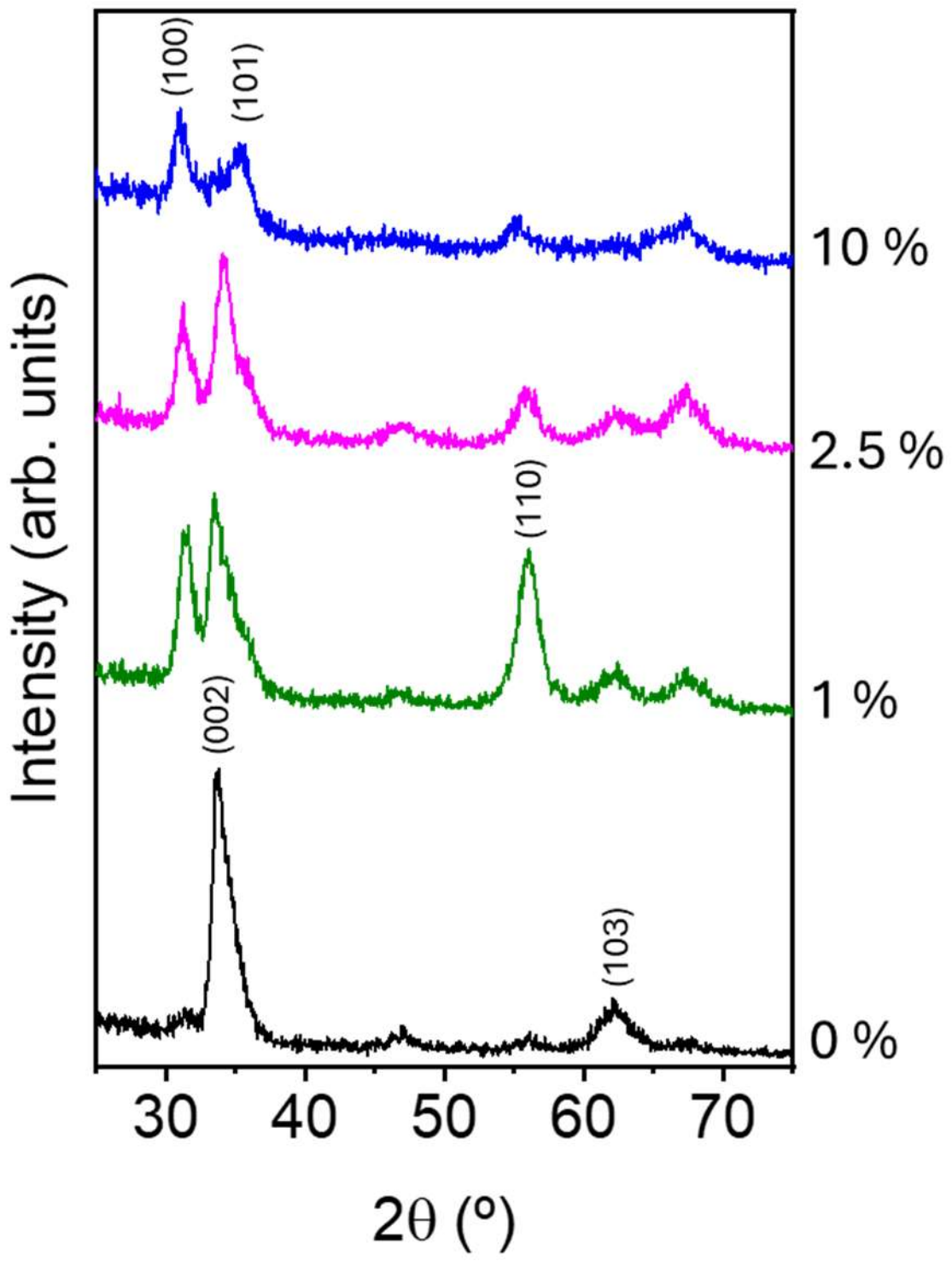

**Figure 3.** Diffractograms of films grown on glass with different ZnO:$Yb_2O_3$ contents. The doping level expressed as wt.% in the PLD target is shown on the right side of the figure. The main diffraction peaks of pure ZnO are identified.

The strong modification of the ZnO crystalline structure for doping levels above 2.5 wt.% is responsible for the observed decrease in the evolution of the carrier density and mobility which can be appreciated both in **Table 1** and in the features of the optical transmission spectra in **Figure 1**. For a nominal doping of 10.0 wt.%. the films remain essentially transparent in the near IR spectral region while the transmission shows only a relatively small diminution for the 5.0 wt.%.

**3.2. Laser processing of ZnO:$Yb^{3+}$ films with fs laser pulses**

Films with nominal doping levels of 1.0 wt. %, 5.0 wt. % and 10.0 wt. % were processed with 350 fs laser pulses at 1030 nm with a repetition rate $f_{rep}$ = 500 kHz, a beam scanning speed of $v$ = 2.5 m/s, and a scan line separation $d$ = 4 µm. Small areas (1 × 1 $mm^2$) were produced with fluences in the 0-800 mJ/$cm^2$ with the beam polarization perpendicular to scan direction to facilitate the formation of coherently propagated LSF-LIPSS [43]. **Figure 4** shows several SEM images of samples with different doping levels irradiated at a fluence $F$ = 590 mJ/$cm^2$.

It can be clearly seen (cf. **Figure 4(a-c)**) that for the three doping levels shown, LSF-LIPSS features are observable in the surface along with distinct ablative features. In contrast, for the undoped film (inset in **Figure 4(d)**), LIPSS formation is not observed in the analyzed fluence interval (up to 1000 mJ/cm$^2$). As soon as the modification threshold of the undoped film is overcome, the film gets deeply ablated and ZnO remains only in the space between the scan lines. This can be appreciated in the optical micrograph (**Figure 4(d)**) because the scan separation was increased to $d$ = 20 µm, close to the beam diameter (Ø = 29.2 µm).

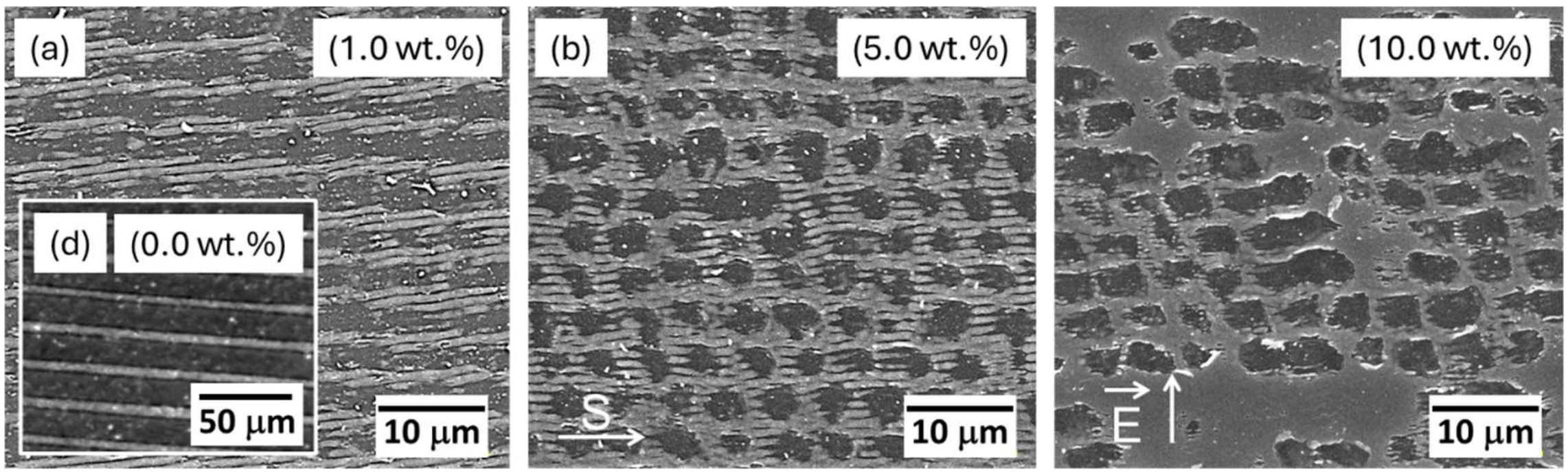

**Figure 4.** (a-c) SEM images at the same magnification of $ZnO:Yb_2O_3$ films with the nominal composition indicated in wt.% irradiated with 590 mJ/cm$^2$ fs-laser pulses at 500 kHz. The scan-line separation in all cases is $d$ = 4 µm. (d) Optical reflection micrograph of an undoped film irradiated at 614 mJ/cm$^2$ with a scan separation of $d$ = 20 µm and a speed $v$ = 1 m/s. S and $\vec{E}$ arrows indicate the orientation of the laser scanning direction and laser beam polarization.

For a doping level of just 1.0 wt.% ($n_e \sim 3 \times 10^{19}$ cm$^{-3}$), LSF-LIPSS with a separation of ~ 0.9 µm, oriented perpendicular to the beam polarization can be clearly appreciated. LSF-LIPSS morphologies are imbricated with ablating features specific of the doping level considered. For the lowest doping value, the surface is ablated in the form of relatively broad stripes (dark contrast) alternating with LIPSS striped regions. The apparent lower "strength" of the ablative features is likely related both to the higher free-electron density and the lower lattice deformation caused by doping. When the doping increases to 5.0 wt.% although the free electron density is very similar, the lattice disorder increases substantially (see **Figure 3**) leading to a situation where each individual pulse ablates the film and LIPSS are observed at the non-ablated edges of the individual laser spots. We can see a quasi-periodic distribution of ablated spots with a vertical separation of 4 µm (which corresponds de the scan line separation, $d$) and a horizontal separation of 5 µm, corresponding to the "landing" of the individual pulses (pulse-to-pulse time separation of 2 ns) at a laser scanning speed of 2.5 m/s. Indeed, such a periodic distribution of ablated spots can be appreciated for the three doping levels shown in **Figure 4**, although is less obvious.

When the doping reaches 10.0 wt.%, the free carrier density decreases down to $9.3 \times 10^{18}$ cm$^{-3}$ and the LSF-LIPSS can be barely appreciated at the edges of the spots landing at the surface. Instead, a nearly homogeneously molten and re-solidified ZnO regions can be appreciated at the surface. Globally, these features allow establishing a lower limit of free carrier density for the formation LSF-LIPSS of ~ $10^{19}$ $cm^{-3}$ and point out to the role of the thermodynamic properties of ZnO in the observed morphologies. ZnO has a very high melting point of 1974 °C and is known to substantially sublimate at high temperatures by dissociation to zinc atoms and oxygen molecules [44]. Because of the rapid heating of the film with fs-laser pulses, it is possible to reach temperatures close to and above the melting point, but which would lead to the rapid thermal decomposition of the regions at the higher temperature. Consequently, at the edges of the overlapping spots, where the local fluence is smaller, LSF-LIPSS can be formed whereas the material in the central region of the spots will both melt and sublimate. In the case of the 10.0 wt.% doping, the diminution of the free carrier density below $10^{19}$ $cm^{-3}$ hampers the formation of LIPSS inducing a relatively homogenous melting and sublimation process. On the other hand, both the absorption of the film, and the thermal conductivity of the substrate must affect morphology as both parameters condition the local peak temperature reached during and after the interaction with the laser, and thus the extent of melt and sublimation occurring at the surface.

To further confirm this hypothesis, samples with a 5 wt. % doping were grown on fused silica substrates at the same $O_2$ pressure ($2 \times 10^{-2}$ mbar) and processed under similar conditions than those of the films shown in **Figure 4**, even slightly increasing the fluence. The achieved results are illustrated in **Figure 5** where we compare the results corresponding to a sample grown in soda-lime glass (**Figure 5(a)**, **(b)**) and in fused silica (**Figure 5(c)**, **(d)**). Despite having similar levels of free carrier density (~ $3\times10^{19}$ $cm^{-3}$) the observed morphologies are very different. As already indicated, in the first case, LSF-LIPSS are observed only at the edges of the individual spots landing on the surface while for the second one, smooth, coherent, long-range propagation [43,45] of LSF-LIPSS can be clearly appreciated. The thermal conductivity of the used soda-lime glass substrate is $\kappa_{glass} = 0.9$ $W\cdot m^{-1}\cdot K^{-1}$ and that of fused silica is $\kappa_{silica} = 1.4$ $W\cdot m^{-1}\cdot K^{-1}$. Therefore, a change of thermal conductivity of approximately 50% enables us to observe extremely high quality, coherently propagated structures with very narrow valleys, apparently very deep. The higher temperature reached at the interference maxima of the LIPSS is enough to very strongly enhance the local material sublimation at the valleys while the increase of the thermal conductivity of the substrates helps in achieving a rapid cooling of the ridges of the structure which remain above the strong sublimation temperature interval for a reduced amount of time.

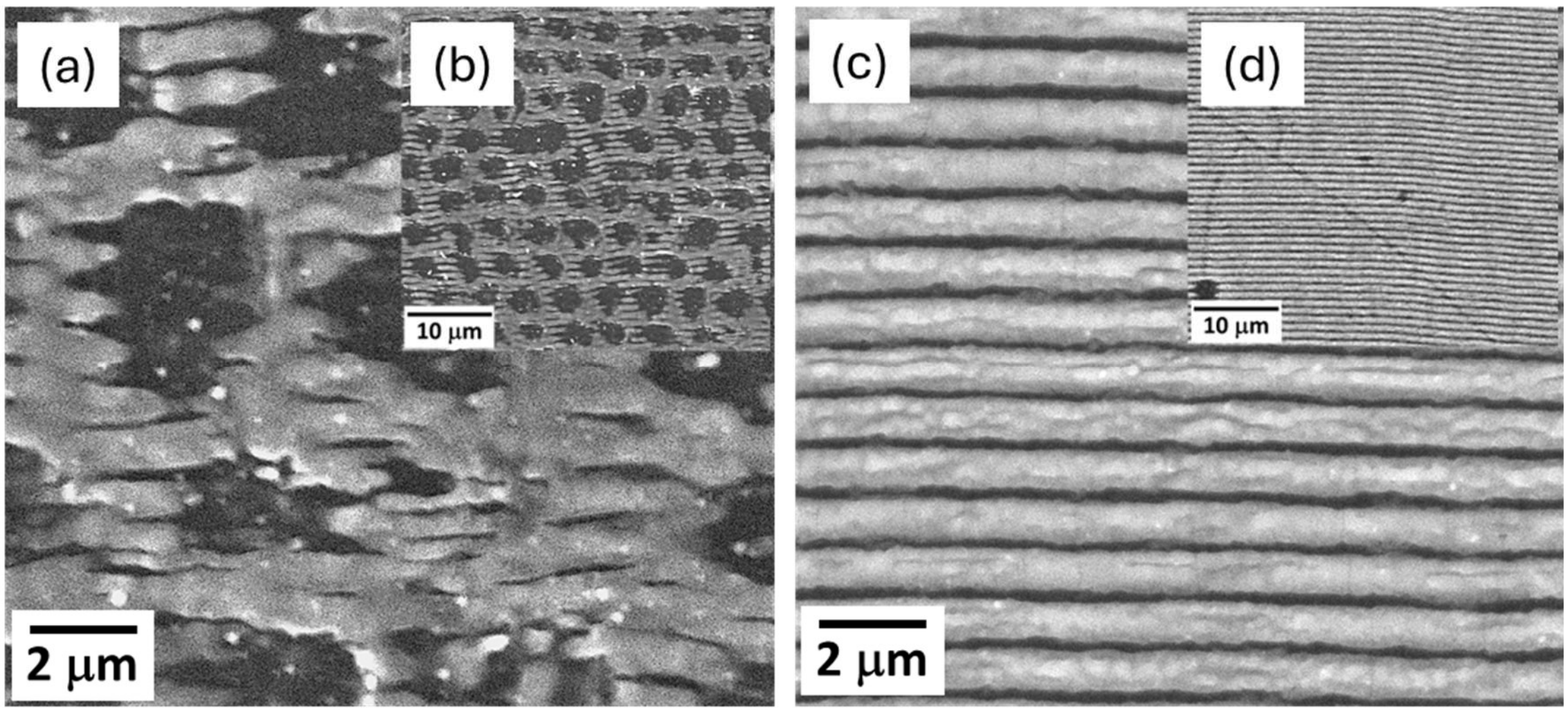

**Figure 5.** SEM images at the same magnification of doped films (5.0 wt.%) deposited on (a, b) soda-lime glass and (c, d) fused silica irradiated with (a, b) 590 mJ/cm$^2$ and (c, d) 618 mJ/cm$^2$ fs-laser pulses at 500 kHz with $d$ = 4 μm and $v$ = 2.5 m/s.

Another remarkable aspect of these results is that the observed structures can be clearly described as coherently propagated LSF-LIPSS while in most of the early studies on LIPSS formation in ZnO under ultrashort laser pulses, mainly high spatial frequency (HSF) LIPSS are observed. Indeed, LIPSS formation mechanisms in ZnO samples has been analyzed in detail by several authors since the early work from Zamfirescu et al. [46]. The role of surface second harmonic generation (SHG) in the observation of HSF-LIPSS in ZnO was pointed out by Guo and coworkers [47,48], who observed a strong emission of SHG radiation during the formation of HSF-LIPSS and other nanostructures in ZnO irradiated with 250 kHz fs-laser pulses at 800 nm. Several authors have observed the formation of HSF-LIPSS or mixtures of LSF- and HSF-LIPSS upon irradiation of ZnO or doped ZnO with low repetition rate fs-laser pulses at 800 nm [6,49]. The importance of surface SHG generation during LIPSS formation was revisited by Dufft and coworkers [50] to include the role of the transient modification of the optical properties of the material upon laser excitation. They concluded that the combination of the "classical LIPSS theory" by Sipe and coworkers [51] with the use of the Drude model to describe the transient properties of the excited material are sufficient to describe why different laser processing parameters lead to the observation of HSF- and LSF-LIPSS in ZnO. Our observations are perfectly consistent with the work by Dufft [50] and with the more recent work by Feng and coworkers [52]. By comparing our results with the ones in Ref. [50], we can see that the use of a high repetition rate, a high scanning speed and, more importantly, the presence of an initial carrier density close to $10^{20}$ $cm^{-3}$ (thus close to the point where the material will show a metallic behavior at 1030 nm) enables the formation of LSF-LIPSS. This feature has also been observed in other transparent materials, such as ITO [3] and FTO [4].

### 3.3. Optimization of LIPSS structures in $ZnO:Yb^{3+}$ films processed with fs laser pulses

To investigate the production of large electrical anisotropies upon coherent propagation of LSF-LIPSS we produced a second set of samples aiming at maximizing their free carrier density while preserving their transparency in the visible spectral region. For this purpose, the $O_2$ dynamic pressure was finely tuned to increase further the free carrier density of the films produced from doped targets.

As indicated in the review by Perrière et al. [40], a clear advantage of PLD for multifunctional ZnO synthesis is the precise control of the oxygen incorporation in films. This enables tuning the electrical conductance over several orders of magnitude. We therefore optimized the oxygen growth pressure using four-probe measurements as a feed-back tool when growing samples from targets of the indicated $Yb_2O_3$ contents at different oxygen pressures. As a result, a new set of optimized samples was grown at a pressure of $7.0 \times 10^{-3}$ mbar on soda-lime glass, fused silica and sapphire substrates. The most relevant electrical parameters of the as deposited films on soda-lime glass are shown in **Table 2**.

**Table 2.** Electrical parameters of $ZnO:Yb_2O_3$ films grown on glass substrates using the indicated targets (the content of $Yb_2O_3$ in wt.% is indicated in the left-most column). Films were deposited at a dynamic $O_2$ pressure of $7 \times 10^{-3}$ mbar in all cases and with a thickness ~ 350 nm.

| $Yb_2O_3$ [wt.%] | Carrier density $n_e$ [$cm^{-3}$] | Carrier mobility [$cm^2/sV$] | Resistivity [$\Omega \times cm$] | Sheet resistance [$\Omega$ / □] |
|---|---|---|---|---|
| 1.0 | $7.62 \times 10^{19}$ | 14.2 | $5.76 \times 10^{-3}$ | $1.61 \times 10^{2}$ |
| 5.0 | $1.94 \times 10^{20}$ | 14.8 | $2.17 \times 10^{-3}$ | $0.61 \times 10^{2}$ |
| 10.0 | $3.55 \times 10^{20}$ | 9.9 | $1.79 \times 10^{-3}$ | $0.50 \times 10^{2}$ |

The carrier density increases linearly with the $Yb_2O_3$ content in the targets($R^2$ ~ 0.999), with a fitted slope of $3.10 \times 10^{19}$ $cm^{-3}$ wt.% and an intercept of $4.30 \times 10^{19}$ $cm^{-3}$ (see SIF, **Figure S3**).

This later value would correspond to the contribution of oxygen vacancies to the film conductivity in the undoped film. It is also remarkable that the presence of oxygen vacancies greatly improves the carrier density and film conductance although for the highest doping level, likely because of the diminution of grain size, the mobility is substantially reduced when compared to the lower doping levels. X-Ray diffraction performed in these samples points also towards the influence of the oxygen partial pressure on the defect structure and hence on the electrical properties (see SIF, **Figure S4**).

The optical transmission of the optimized films is shown in **Figure 6**. For the three doping levels, the optical behavior is like that of typical TCO´s like ITO or FTO [3,4] with a broad transparent region up to ~ 1.2 μm followed by a transmission decay caused by the presence of free carriers at room temperature with densities above $10^{19}$ $cm^{-3}$, as shown in **Table 2**. These films were irradiated with fs-laser pulses with the same parameters already indicated in **Section 2.2** using local fluences in the 0-800 $mJ/cm^2$.

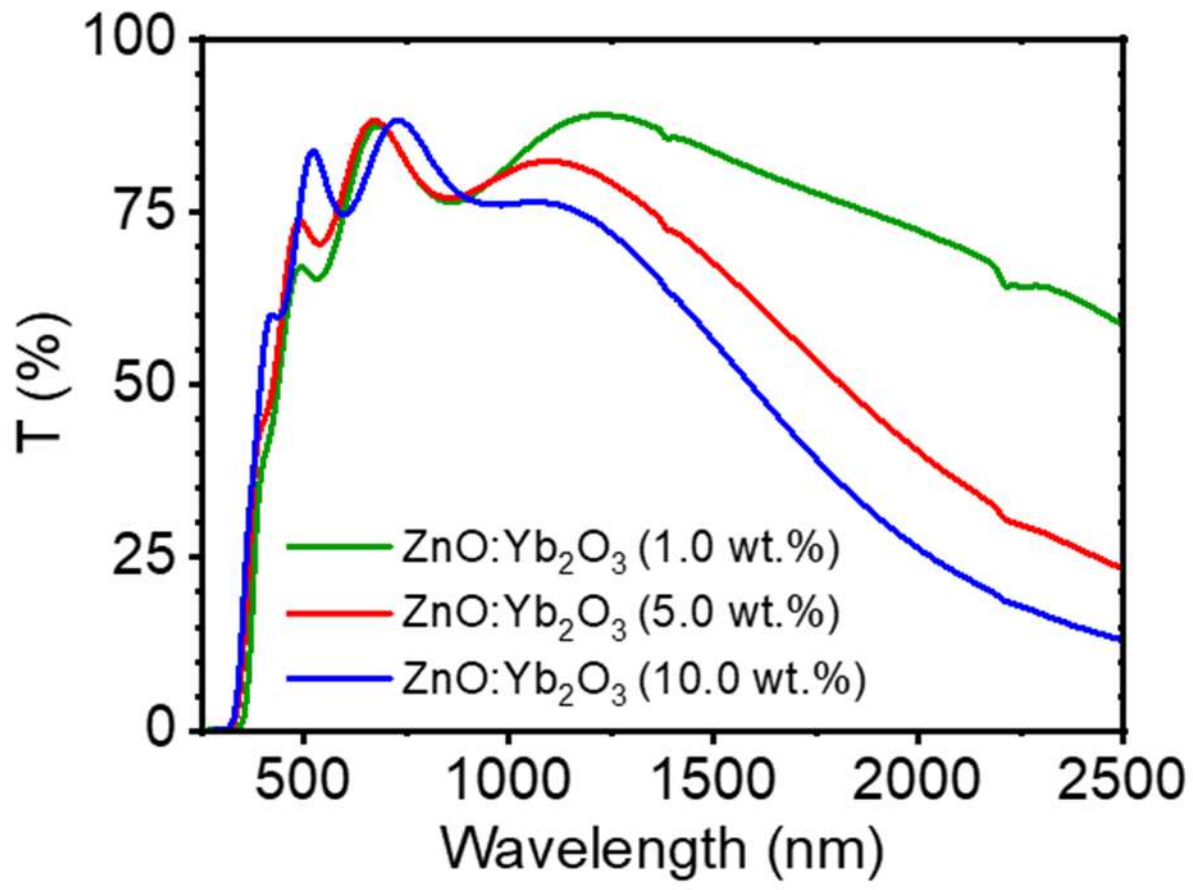

**Figure 6.** Optical transmission spectra of films grown on fused silica substrates from targets with different $ZnO:Yb_2O_3$ contents (indicated as wt.%) at an $O_2$ pressure of $7.0 \times 10^{-3}$ mbar. The film thickness was to ~ 350 nm in all cases.

**Figure 7** shows illustrative images of the structures generated in the optimized thin film samples deposited on fused silica substrates using two different pulse durations. Despite the clear regularity of the observed LSF-LIPSS for a pulse duration of 350 fs, small cracks are observed at the ridges of the LIPSS. This impedes preserving the macroscopic electrical conductivity along the LIPSS direction and makes it impossible to induce a macroscopic electrical anisotropy in the laser processed surfaces. Additionally, we can see that in the space between the ridges (black contrasted) the film seems completely removed, indicating that the valleys of the structure, with a width around 250 nm, reach the substrate. This later feature is confirmed by the cross-sectional image in **Figure 7(b)** where we can appreciate other important features. LIPSS are highly regular with a nearly hemispherical shape that indeed reaches the substrate. Additionally, the maximum height of the structures at the ridges is like that of the pristine film, indicating that the sublimation process occurs mainly at the interference maxima of the LIPSS.

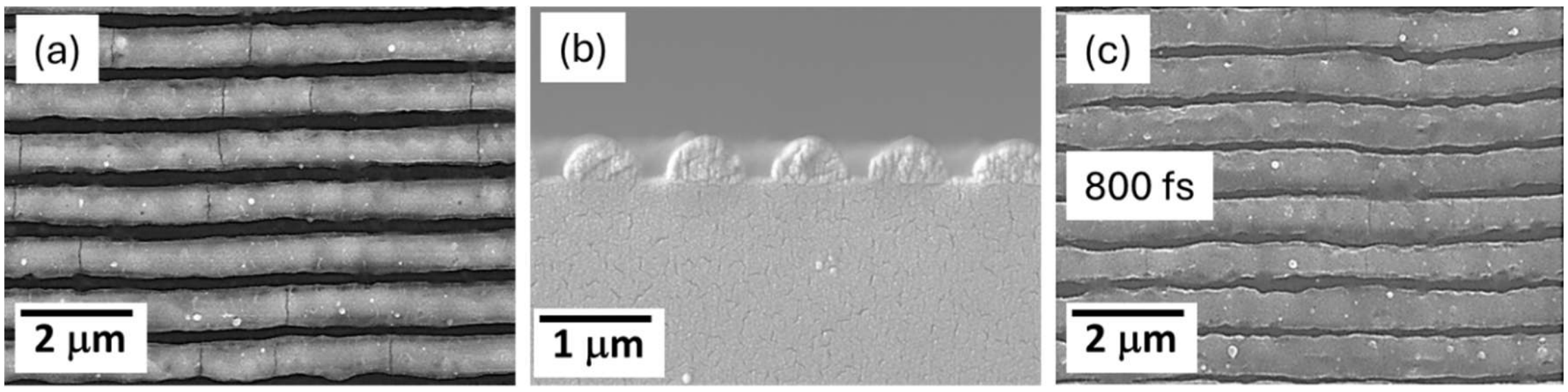

**Figure 7.** (a) SEM image of doped (5.0 wt.%) films on fused silica irradiated with 350 fs-laser pulses at 471 mJ/cm$^2$ and (c) 800 fs fs-laser pulses at 621 mJ/cm$^2$. In all cases, $f_{rep}$ = 500 kHz, $d$ = 4 μm and $v$ = 2.5 m/s. The central image (b) shows a cross-sectional image of the sample in (a). Notice that the fine structure observable in the substrate region in (b) is caused by the Au-film deposited to avoid charge effects in the cross-sectional image at the SEM.

The comparison of these features to ones in **Figure 5(a, b)** confirms that the generated morphologies can be controlled via substrate conductivity. As a reference, in the SIF, in **Figure S5**, a comparison of the different substrates used under the exact same laser processing conditions is shown. This also makes clear that the local temperature evolution at the irradiated regions over time plays a key role in defining the shape of the LSP-LIPSS. This was also confirmed by processing a sample with 800 fs laser pulses to make the avalanche process occurring upon pulse absorption to be smoother. The result is shown in **Figure 7(c)** where similarly high quality, coherently propagated LSF-LIPSS are evident. However, and more importantly, the use of a longer pulse duration, even for higher fluences, suppresses the formation of deep cracks at the ridges of the structure preserving, at least partly, the macroscopic conductivity of structure along the LIPSS direction.

The feasibility of controlling the LSF-LIPSS morphology via thermal properties of the film/substrate ensemble is even clearer by comparing the results achieved in samples on fused silica ($\kappa_{silica}$= 1.4 W·m$^{-1}$·K$^{-1}$) with those observed in samples grown on sapphire ($\kappa_{sapphire}$ ~ 10 Wm$^{-1}$K$^{-1}$ at high temperature), as illustrated in **Figure 8**. For low fluences, HSF-LIPSS poorly propagated are induced (**Figure 8(a)**), consistently with the observations in Refs. [50,52].

Over a threshold (~ 460 mJ/cm$^2$), highly regular coherent propagation is induced with narrow valleys and ridges apparently flat (**Figure 8(b)**). A few droplets can be appreciated on the surface but unlike the case of films on fused silica (see **Figure 7 (b)**) cracking of the structures is not observed even at the largest fluences used (**Figure 8(c)**), which is above the fluences used till this moment in the presented data. Interestingly, EDX measurements in the pristine and laser irradiated samples both on silica and sapphire substrates show that the ratio Yb/Zn (at.%) remains unchanged within experimental error until the laser fluence increases beyond 590 mJ/cm$^2$ when it starts to increase, as expected from the much higher mass of $Yb^{3+}$ ions. This result is extremely interesting as it indicates that the composition of the material remaining at the ridges is not substantially modified upon LSF-LIPSS formation. This is a situation very different than that observed in ITO and FTO upon coherent propagation of LSF-LIPSS irradiated at 1030 nm [3,4] where the LIPSS formation mechanisms are related to the compositional changes induced upon multi-pulse laser exposure.

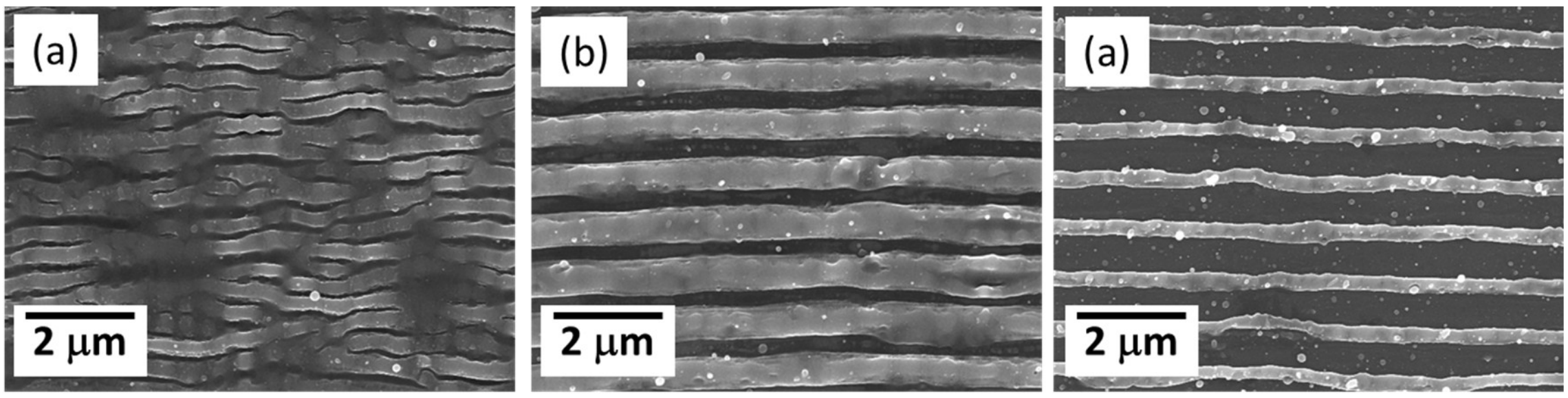

**Figure 8.** (a-c) SEM images at the same magnification of films doped films (5.0 wt.%) on sapphire irradiated with (a) 441, (b) 489 and (c) 671 mJ/cm$^2$ fs-laser pulses at 500 kHz with $d$ = 4μm and $v$ = 2.5 m/s for a pulse duration of 350 fs.

The resistivity of the laser processed films on sapphire was measured in large areas (6 × 6 mm$^2$) for different fluences in the studied range. Just above the fluence threshold to form LSF-LIPSS, areas processed at 475 mJ/cm$^2$ show an anisotropy factor $\xi \sim 2$, calculated as $\xi = R_{\square\perp}/R_{\square\parallel}$, while preserving the sheet resistance measured in the direction parallel to the LIPSS ($R_{\square\parallel}$) under 300 $\Omega\cdot\square^{-1}$. By increasing the processing fluence up to 540 mJ/cm$^2$, the anisotropy factor reaches values as high as $\xi = 89$, but then $R_{\square\parallel} > 150$ k$\Omega\cdot\square^{-1}$, making it impossible to integrate the laser-processed TCO into a real device. In between these two laser fluence, at 500 mJ/cm$^2$, the sheet resistance values achieved are $R_{\square\parallel} = 920$ $\Omega\cdot\square^{-1}$ in the parallel direction and $R_{\square\perp} = 7.5$ k$\Omega\cdot\square^{-1}$ in the perpendicular one, providing an anisotropy factor $\xi \sim 8$. At this intermediate fluence, $R_{\square\parallel}$ is only a factor 16 compared to that of the pristine material, which can be explained by the cross-sectional area diminution of the LIPSS regions.

These results globally show that the generation and coherent propagation of LSF-LIPSS does compromise the electrical properties of the material remaining at the ridges and suggests that large area applications of the LIPSS-processed material are feasible.

## 4. Conclusion

As a general conclusion, we have demonstrated the key role of free carrier density in the formation and coherent propagation of LSF-LIPSS. By using ad-hoc doped ZnO as a paradigm, we have shown that, above $n_e = 10^{19}$ $cm^{-3}$, LSF-LIPSS can be successfully produced and propagated over macroscopic regions at large processing speed using NIR (1030 nm) fs-laser pulses at high repetition rate (100´s of kHz). The quality of the microstructures generated in the case of ZnO are conditioned by the thermal conductivity of the substrate and the particular pulse duration used. Despite this, we have been able to produce electrically anisotropic surfaces with conductivities along the LIPSS about ten times higher than perpendicularly to them and with a relatively small penalty in terms of the electrical properties of the material remaining at the ridges. This enables the use of LIPSS-processed surfaces for a variety of macroscopic applications both using doped $ZnO:Yb^{+3}$ and other ZnO-based TCOs.

## 5. Acknowledgements

This research was partially funded by the Spanish Ministry of Science, Innovation and Universities (MICIU), the Spanish Research Agency (AEI- MICIU), and the European Regional Development Fund (ERDF) through grant numbers PID2023-148178OB-C22 (HyperSpec) and PID2024-156974OB-C21 (SLIM-2P), funded by MCIN/AEI/10.13039/501100011033 and by 'ERDF (FEDER) A way of making Europe'. This research was also partially funded by the Consejería de Educación, Ciencia y Universidades de la Comunidad de Madrid through grant number TEC-2024/TEC-85 (MATRIX-CM). This work has been also supported by the MaX-CSIC Excellence Award SYNERGY-IO/03 (PhotoLaP). RA acknowledges the funding from a Marie Skłodowska-Curie Postdoctoral grant (Nº 101149132). The help of M. I. Maher for his assistance in SEM, and EDX measurements performed at the facilities of CENIM-CSIC (Madrid, Spain) is similarly greatly appreciated.

## 6. Author contribution list

**Gonzalo Gomez-Munoz:** Conceptualization, Data Curation, Investigation, Methodology, Writing original draft, Writing -review & editing. **Rocio Ariza:** Investigation, Data Curation, Writing -review & editing. **J.F. Ramos-Justicia:** Investigation, Writing -review & editing

**Fátima Cabello-Pardos:** Data Curation, Investigation, Writing -review & editing. **Antonio Pablo Perez:** Investigation, Writing -review & editing. **Jose Gonzalo:** Investigation, Methodology, Funding Acquisition, Project administration, Supervision, Writing -review & editing. **Belen Sotillo:** Investigation, Writing -review & editing. **Binbing Xia:** Investigation, Writing -review & editing. **Paloma Fernandez:** Funding Acquisition, Investigation, Methodology, Project administration, Supervision, Writing -review & editing. **Carmen Lopez-Santos:** Investigation, Methodology, Writing -review & editing. **Javier Solis:** Conceptualization, Funding Acquisition, Investigation, Methodology, Project administration, Supervision, Writing original draft, Writing -review & editing. All authors reviewed and approved the final manuscript.